\documentclass[twocolumn,twoside,5p,preprint]{elsarticle}
\biboptions{sort&compress}

\usepackage[utf8]{inputenc}
\usepackage[english]{babel}
\usepackage{amssymb,amsthm,amsmath,amstext,amsbsy,amsopn}
\usepackage{bbm}
\usepackage{graphicx}
\usepackage{isotope}
\usepackage{float}
\usepackage[dvipsnames]{xcolor}
\usepackage[T1]{fontenc}

\usepackage{hyperref}
\hypersetup{
    colorlinks = true,
    citecolor  = blue,
    linkcolor  = blue,
    urlcolor  = black
}

\newcommand{\MeV}{\text{MeV}}

\newcommand{\ph}[1]{{$#1$}p{$#1$}h}
\newcommand{\ai}{{\emph{ab initio}}}
\newcommand{\hilbert}[1]{\mathcal{H}_{#1}}

\newcommand{\Xmax}[1]{#1_\text{max}}

\newcommand{\NN}{\text{NN}}
\newcommand{\NNN}{\text{3N}}

\newcommand{\elem}[2]{$^{#2}\text{#1}$}
\newcommand{\Etwop}{E^\star_{2^+}}

\begin{document}

\title{Combining the in-medium similarity renormalization group with the\\ density matrix renormalization group: Shell structure and information entropy}

\address[tud]{Technische Universit\"at Darmstadt, Department of Physics, 64289 Darmstadt, Germany}
\address[emmi]{ExtreMe Matter Institute EMMI, GSI Helmholtzzentrum f\"ur Schwerionenforschung GmbH, 64291 Darmstadt, Germany}
\address[mpik]{Max-Planck-Institut f\"ur Kernphysik, Saupfercheckweg 1, 69117 Heidelberg, Germany}
\address[algo]{Algorithmiq Ltd., Kanavakatu 3C, FI-00160 Helsinki, Finland}
\address[she]{SHE Chemistry, GSI Helmholtz Centre for Heavy Ion Research, Planckstr.~1, 64291 Darmstadt, Germany}
\address[eth]{ETH Z\"urich, Laboratory for Physical Chemistry, Vladimir-Prelog-Weg 2, 8093 Z\"urich, Switzerland}
\address[atk]{Institute for Nuclear Physics, P.O. Box 51, H-4001 Debrecen, Hungary}
\address[wigner]{Wigner Research Centre for Physics, P.O. Box 49, 1525 Budapest, Hungary}
\address[ias]{Institute for Advanced Study, Technical University of Munich, Lichtenbergstrasse 2a, 85748 Garching, Germany}
\address[dtpbudapest]{Department of Theoretical Physics, Institute of Physics, Budapest University of Technology and Economics, M\H uegyetem rkp 3., H-1111 Budapest, Hungary}
\address[oradea]{Department  of  Physics,  University  of Oradea,  410087,  Oradea,  Romania}
\address[mtabme]{MTA-BME Quantum Dynamics and Correlations Research Group, Budapest University of Technology and Economics, M\H uegyetem rkp. 3., H-1111 Budapest, Hungary}

\author[tud,emmi,mpik]{A.\ Tichai}
\ead{alexander.tichai@physik.tu-darmstadt.de}

\author[algo,she,eth]{S.\ Knecht}
\ead{stefan@algorithmiq.fi}

\author[atk]{A.\ T.\ Kruppa}
\ead{atk@atomki.hu}

\author[wigner,ias]{\"O.\ Legeza}
\ead{legeza.ors@wigner.hu}

\author[dtpbudapest,oradea]{C.~P.~Moca}
\ead{mocap@uoradea.ro}

\author[tud,emmi,mpik]{A.\ Schwenk}
\ead{schwenk@physik.tu-darmstadt.de}

\author[mtabme,dtpbudapest,wigner]{M.\ A.\ Werner}
\ead{werner.miklos@ttk.bme.hu}

\author[mtabme,dtpbudapest]{G.\ Zarand}
\ead{zarand.gergely.attila@ttk.bme.hu}

\begin{abstract}
We propose a novel many-body framework combining the density matrix renormalization group (DMRG) with the valence-space (VS) formulation of the in-medium similarity renormalization group.
This hybrid scheme admits for favorable computational scaling in large-space calculations compared to direct diagonalization.
The capacity of the VS-DMRG approach is highlighted in \ai{} calculations of neutron-rich nickel isotopes based on chiral two- and three-nucleon interactions, and allows us to perform converged \ai{} computations of ground and excited state energies.
We also study orbital entanglement in the VS-DMRG, and investigate 
nuclear correlation effects in oxygen, neon, and magnesium isotopes. The explored entanglement measures reveal nuclear shell closures as well as pairing correlations.
\end{abstract}

\maketitle

\section{Introduction}
Low-energy nuclear theory has seen dramatic progress in the \ai{} description of nuclei based on chiral effective field theory (EFT) interactions and powerful many-body methods that can access nuclei up to $^{208}$Pb~\cite{Herg20review,Hebe203NF,Stroberg2021,Hu2021lead}. The use of EFTs enables consistent two- and many-body interactions (and operators)~\cite{Epel09RMP,Mach11PR,Hamm13RMP,Bacca14JPG,Hamm20RMP} as well as theoretical uncertainty estimates from the EFT power-counting expansion~\cite{Epel15improved,Furn15uncert,Weso213N}.
In medium-mass nuclei the many-body Schr\"odinger equation is commonly solved using basis expansion methods that incorporate low-rank particle-hole excitations in a systematic way from a suitably chosen reference state~\cite{Hage14RPP,Herg16PR,Soma20SCGF,Tichai2020review}.
The development of methods with polynomial computational scaling was key for advancing \ai{} calculations to heavier systems~\cite{Arthuis2020a,Sun21CCEI,Stroberg2021,Miyagi2021,Hu2021lead}.
Among these, the in-medium similarity renormalization group (IMSRG) represents a powerful and flexible approach to efficiently target a broad range of nuclear observables~\cite{Tsuk11IMSRG,Herg16PR,Stro17ENO,Gebr17IMNCSM,Stroberg2019,Yao2020,Heinz2020}.
However, the description of strongly correlated, open-shell systems poses significant challenges when a (symmetry-conserving) single-reference state does not capture the static correlations.
This requires the development of novel expansion schemes at tractable computational cost~\cite{Herg20review}. In this work, we use valence-space (VS) techniques where an active-space Hamiltonian is decoupled from a closed-shell core and subsequently used in a large-space or shell-model diagonalization, giving access to a wide range of observables such as low-lying spectroscopy and transitions~\cite{Holt14Ca,Bogn14SM,Jans14SM,Stro17ENO,Sun2018,Stroberg2019}.
In \ai{} calculations, open-shell systems are also targeted using either symmetry-broken~\cite{Soma13GGF2N,Sign14BogCC,Tichai18BMBPT,Novario2020a,Yuan2022,Hagen2022PCC} or multi-configurational reference states~\cite{Herg13MR,Tich17NCSM-MCPT,Gebr17IMNCSM,Frosini2021mrII,Frosini2021mrIII}.

In other fields of many-body research such as condensed matter physics or quantum chemistry, the density matrix renormalization group (DMRG) is well established as a powerful tool to treat strongly correlated quantum systems~\cite{White1992,Schollwoeck2011,szalay2015tensor,baia20a}. Previous studies in nuclear structure have focused on phenomenological shell-model applications~\cite{Papenbrock2005,Pittel2006,Thakur2008jdmrg,Legeza2015} and open quantum systems using a Gamow basis~\cite{Papdimitriou2013gamowdmrg,Foss17TetraN}.
However, the development of the DMRG to medium-mass \ai{} calculations has not been explored. This is the goal of this work.

In this Letter, we apply the DMRG approach in \ai{} nuclear structure calculations of medium-mass nuclei for the first time. We use the VS-IMSRG to decouple a valence-space Hamiltonian, which is then used as input to large-scale DMRG calculations.
The favorable scaling of the DMRG provides an efficient framework for accessing computationally challenging open-shell nuclei in a systematically controllable way.
Moreover, entanglement properties of many-body system are accessible from orbital entropies and derived quantities, thus proving a novel perspective to the emergence of structure from nuclear forces.

\section{Valence-space DMRG approach}
The central idea of this work is the combination of the DMRG with the valence-space formulation of the IMSRG. This gives rise to a hybrid many-body framework, which we refer to as valence-space density matrix renormalization group (VS-DMRG).
Starting from an initial Hamiltonian with two-nucleon (\NN{}) and three-nucleon (\NNN{}) interactions, the VS-IMSRG generates a valence-space-decoupled Hamiltonian that is restricted to an active space of limited size~\cite{Stro17ENO,Stroberg2019}.
While the use of a valence-space Hamiltonian is similar to the phenomenological shell model, with the VS-IMSRG this is derived from chiral EFT interactions without adjustments.
During the IMSRG-evolution many-body operators of higher particle rank are truncated at the normal-ordered two-body level, defining the IMSRG(2) truncation.
The valence-space-decoupled Hamiltonian $H_\text{VS}$ used as input for the DMRG calculation is represented in second-quantized form as
\begin{align}
    H_\text{VS} = 
    \sum_{p} \varepsilon_p \, c^\dagger_p c_p +
    \frac{1}{4} \sum_{pqrs} V_{pqrs} \, c^\dagger_p c^\dagger_q c_s c_r \, ,
\end{align}
where $\varepsilon_p$ are the single-particle energies and $V_{pqrs}$ the (anti-symmetrized) two-body matrix elements.
The collective label $p = (n_p, l_p, j_p, m_p, t_p)$ gathers all quantum numbers of a single nucleon: radial quantum number $n$, orbital angular momentum $l$, total angular momentum $j$ and its projection $m$, and isospin projection $t$ distinguishing protons and neutrons.

The initial VS-IMSRG decoupling is performed in a single-particle space of 15 major harmonic-oscillator shells, i.e., $\Xmax{e} \equiv (2n +l)_\text{max} = 14$, and the \NNN{} interaction matrix elements are restricted to $e_1 + e_2 + e_3 \leqslant E_{3\text{max}} = 16$.
For all our calculations, we employ the 1.8/2.0 NN+3N Hamiltonian from Ref.~\cite{Hebe11fits}, which is based on chiral EFT interactions.
The three-nucleon interactions are taken into account by keeping only two-body contributions after normal ordering~\cite{Hage07CC3N,Roth12NCSMCC3N,Herg13IMSRG}.

In the DMRG calculation we use the occupation-number representation of an orbital, yielding a local Hilbert space with dimension $d=2$. Therefore, each orbital is represented by two distinct occupation states $\sigma$, i.e.  $\sigma\in \{0,1\}$. 
The full Hilbert space of $N$ orbitals is then built from a tensor product of the local spaces, i.e., $\hilbert{}^{N} \equiv \otimes_{i=1}^N \hilbert{i}$.
The DMRG approach provides a variational procedure for the minimization of the ground-state energy (or the lowest energy for a given total angular momentum and parity) using a matrix product state (MPS) parametrization of the many-body state (see, e.g., Ref.~\cite{Schollwoeck2011}), that eventually converges to the full configuration interaction (FCI) limit for a given Hilbert space.
To this end, the nuclear orbitals are mapped onto a one-dimensional chain.
This protocol is based on the two-orbital mutual information (see next section) of the orbitals to minimize long-range correlations, i.e., to find a quasi-optimal ordering of the orbitals along the one dimensional DMRG topology~\cite{Rissler2006,Barcza-2011}.

The corresponding wave function of $N$ orbitals is an $N$ dimensional tensor, the CI coefficient corresponding to a determinant  $\boldsymbol{\sigma}=(\sigma_1,\sigma_2,\ldots, \sigma_i,\sigma_{i+1},\ldots,\sigma_N)$ is expressed as a product of matrices $A_i^{\sigma_i}$ associated to each orbital $i$ as  $| \Psi \rangle = \sum_{\boldsymbol{\sigma}}C_{\boldsymbol{\sigma}}|\boldsymbol{\sigma}\rangle$, 
where
\begin{align}
   C_{\boldsymbol{\sigma}} = A_1^{\sigma_1} A_2^{\sigma_2} \ldots A_i^{\sigma_i} A_{i+1}^{\sigma_{i+1}} \ldots A_N^{\sigma_N}\,.
    \label{eq:mps}
\end{align}
The dimension of the matrices in the MPS representation scales exponentially with the number of orbitals, such that truncations are required to keep the the dimensions numerically tractable.
In the DMRG algorithm the matrices $A_i^{\sigma_i}$ are iteratively optimized. In an iteration step of 
the two-site DMRG variant the tensor space is split according to $\hilbert{}^{N} = \hilbert{}^{(\text{left})} \,\otimes\, \hilbert{p} \,\otimes\, \hilbert{p+1} \,\otimes\, \hilbert{}^{(\text{right})}$ where $\hilbert{}^{(\text{left})}$ ($\hilbert{}^{(\text{right})}$) denote the left (right) blocks 
 that are formed from precontracted $A$ matrices to the left and right of the sites $p$ and $p+1$, respectively. 

For a given site $p$, the MPS matrix is updated through a diagonalization of the neighboring block Hamiltonian and the maximal matrix dimension ($M$), also known as bond dimension, is kept below a threshold value by keeping only those matrix components which correspond to highest Schmidt weights obtained via singular value decomposition. 
Therefore, the state's components are obtained through a series of unitary transformations (``sweeps'') going through the orbital space forward from left to right, and then backward, until convergence is reached.
The method's intrinsic truncation error is thus set by $M = \dim \hilbert{}^{(\text{left)}} = \dim \hilbert{}^{(\text{right})}$ corresponding to the dimension of the left/right blocks.
Eventually, the size of the bond dimension to reach an acceptable convergence is in direct correspondence with the amount of quantum entanglement in the many-body state~\cite{szalay2015tensor}.
The DMRG convergence is substantially improved following the configuration-interaction dynamically extended active-space procedure, similar to the calculations performed in Ref.~\cite{Legeza2015}.

\begin{figure}[t]
    \centering
    \includegraphics[clip=,width=0.9\columnwidth]{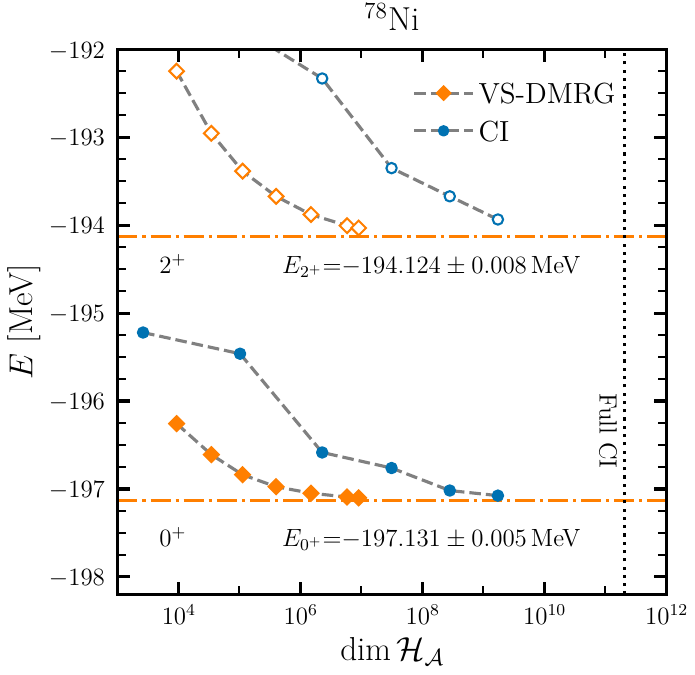}
    \caption{Energies of the ground state (filled) and first $2^+$ state (open symbols) in \elem{Ni}{78} calculated from the VS-DMRG and CI for corresponding many-body-space dimension. The VS-DMRG results are obtained for bond dimensions $M=256,512,1024,2048,4096,8192,10240$, while CI employs a particle-hole truncation with $\Xmax{T}=3,4,5,6,7$. The full valence-space dimension is given by the vertical line (Full CI). Numbers (horizontal lines) correspond to extrapolated VS-DMRG energies.}
    \label{fig:ni78conv}
\end{figure}

\section{Entanglement and correlation measures}
For the study of correlation effects in nuclear many-body systems, we explore a set of entanglement measures~\cite{Vidal2002,Shi2003}.
The total entropy~\cite{Legeza-2004b} $I_\text{tot} \equiv \sum_{p} s_{p}$
is obtained from the single-orbital entropy $s_p \equiv - \text{Tr} \, \rho_p \ln \rho_p$, where $\rho_p$ is the one-orbital-reduced density matrix of the orbital $p$ obtained by tracing out all other orbitals except for $p$~\cite{Legeza2003}.
The single-orbital entropy is directly linked to the natural occupation numbers in the many-body state~\cite{Boguslawski2014orbent}.
Therefore, systems with strong static correlations give rise to increased values for $s_{p}$ and, consequently, $I_\text{tot}$. 
In the case of weakly correlated systems, occupation numbers are either $n_p \approx 0 \, \text{or} \,1$, reflecting the existence of a dominant reference determinant, as obtained in a mean-field calculation, for example.
As a consequence, nuclei with shell closures will be accompanied by a local minimum in the total entropy.
To more cleanly disentangle correlations for protons and neutrons, we define the proton (neutron) total entropy $I_\text{tot}^{(p)}$ ($I_\text{tot}^{(n)}$) where only single-orbital entropies of a given particle species are summed over.
Correlations among pairs of orbitals can be further studied from the entanglement entropy $s_{pq} \equiv -\text{Tr} \, \rho_{pq} \ln \rho_{pq}$ using the two-orbital reduced density matrix $\rho_{pq}$.
Combing single- and two-orbital entropies leads to the mutual information, $I_{p\ne q} \equiv s_p + s_q - s_{pq}$~\cite{Rissler2006}. 
Since matrix elements of $\rho_{pq}$ are expressed in terms of two-orbital correlation functions, also known as generalized correlation functions \cite{Barcza-2015}, $s_{pq}$ can be viewed as a weighted average of the corresponding correlations. Subtraction of $s_p$ and $s_q$ when $I_{pq}$ is calculated is analogous to the usual subtraction of the unconnected parts of the two-orbital correlation functions.
Entanglement studies in nuclear theory have been performed in shell-model applications~\cite{Legeza2015,Kruppa2021} and in no-core calculations of light systems~\cite{Robin2021}.
We emphasize that the entanglement measures are of non-observable character, as they depend on the nuclear Hamiltonian and the many-body basis (see, e.g., Refs.~\cite{Furnstahl02occ,Furnstahl2010}). Thus, we focus on their qualitative behavior.

\begin{figure}[t]
    \centering
    \includegraphics[clip=,width=0.8\columnwidth]{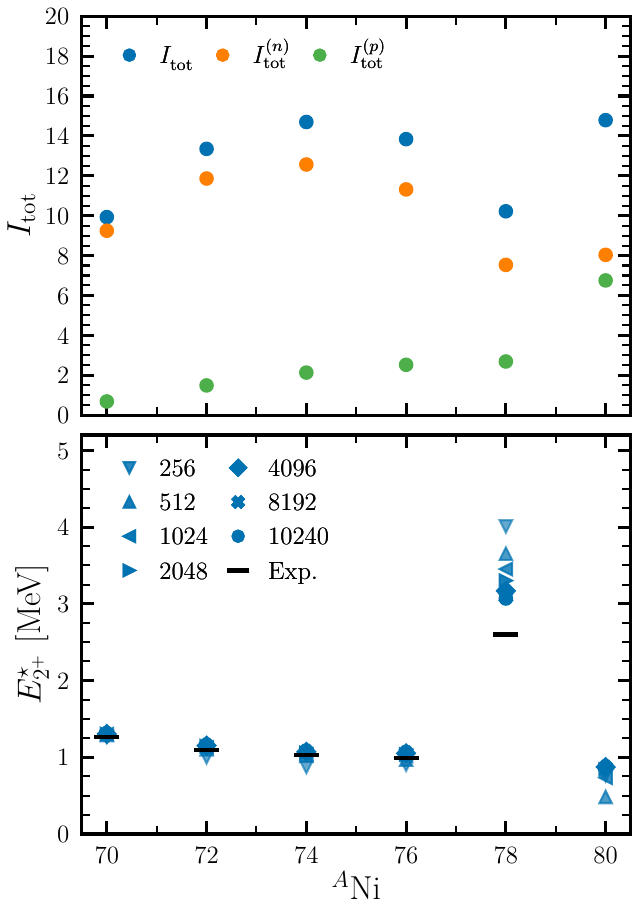}
    \caption{Neutron, proton, and total entropies (top) and $2^+$ excitation energies (bottom) along even-mass nickel isotopes. Entropies are calculated at bond dimension $M=10240$ whereas for the excitation energies the bond dimension was varied between $M=256-10240$. Experimental values are taken from Ref.~\cite{NNDC2022}.}
    \label{fig:nickelshell}
\end{figure}

\section{Neutron-rich nickel isotopes from VS-DMRG}
To show the power of the VS-DMRG, we apply this new approach to the description of neutron-rich nickel isotopes that are attracting significant experimental attention, e.g., with the recent discovery of the doubly magic nature of \elem{Ni}{78}~\cite{Taniuchi2019}.
In fact, \ai{} calculations approaching \elem{Ni}{78}
require additional truncations of the configuration interaction (CI) or shell model space when exploring a $0\hbar \omega$ valence space on top of a $^{60}$Ca core~\footnote{Reference~\cite{Taniuchi2019} quotes the $2^+$ energy for \elem{Ni}{78} to be $\Etwop=3.34\,\MeV$ for including up to $T_\text{max}=7$ particle-hole excitations. In our studies we confirmed that this was a misprint and calculations were performed up to $T_\text{max}=6$.}.
In this work, the CI calculations haven been performed using the \texttt{KSHELL}~\cite{kshell} and \texttt{BIGSTICK}~\cite{Johnson2018bigstick} codes, while the DMRG calculations together with quantum-information-based analysis tools used the \texttt{DMRG-Budapest} program package~\cite{dmrg-budapest}.

\begin{figure}[t]
    \centering
    \includegraphics[clip=,width=\columnwidth]{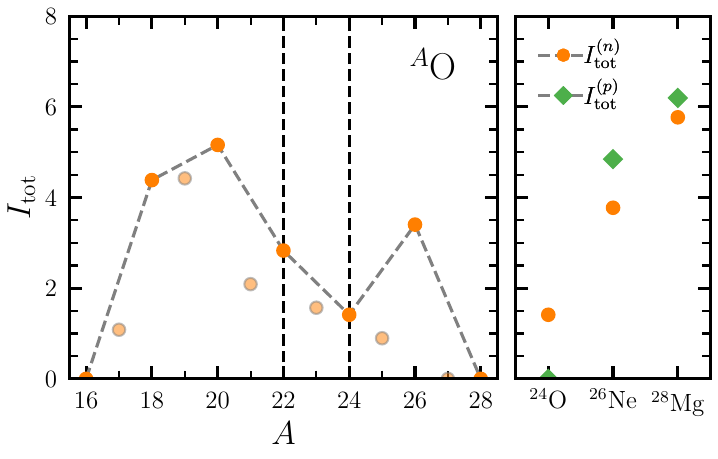}
    \caption{Neutron and proton entropies from VS-DMRG calculations for the oxygen chain (left) and for the evolution at $N=16$ from the closed proton shell to \elem{Ne}{26} and \elem{Mg}{28} (right). Vertical dashed lines indicate neutron shell closures. Total entropies of odd-mass nuclei are displayed as lighter symbols as they are for $M_J=1/2$.}
    \label{fig:ItotONeMg}
\end{figure}

In Fig.~\ref{fig:ni78conv} we compare large-scale CI and VS-DMRG calculations for \elem{Ni}{78} based on the same VS-IMSRG interaction as in Ref.~\cite{Taniuchi2019}.
For \elem{Ni}{78}, the FCI dimension is $2.3 \cdot 10^{11}$, while our largest CI calculations involved $1.9\cdot 10^9$ configurations employing a truncation at $\Xmax{T}=7$ particle-hole (ph) excitations.
In contrast, the dimension of the DMRG space increases only gradually, and is well tractable even for the largest considered bond dimension $M=10240$, with corresponding configuration space of $\approx 10^7$, two orders of magnitude below the largest accessible CI dimension. The DMRG dimension is essentially the dimension of the space spanned by the two block spaces and the two orbitals, $\sim M^2d^2$, further constrained by selection rules for parity, isospin and angular-momentum projection. 
Figure~\ref{fig:ni78conv} clearly shows that the VS-DMRG results for the ground and first $2^+$ excited states reveal a more robust convergence pattern compared to the CI calculation.
While the ground-state energy converges systematically in the CI case, there is still a sizeable linear trend present for the first excited $2^+$ state, making the extrapolation of the excitation energy challenging. 
This may potentially hint at relevant \ph{8} excitations missing in the $\Xmax{T}=7$ truncation.
In contrast, the VS-DMRG results converge systematically beyond $M=1024$.
Fitting a quadratic polynomial $f_\text{extr.}(1/M)=a/M^2 +b/M +c$ enables a robust extrapolation of the energies~\cite{szalay2015tensor}.
Other sweep-based and truncation error based extrapolation procedures have been successfully applied in condensed-matter and quantum chemistry applications~\cite{szalay2015tensor,baia20a,Barcza2022}.
Extrapolation uncertainties are obtained by taking into account only the $3,4,5$ data points corresponding to the largest bond dimensions, yielding a VS-DMRG estimate of $\Etwop = 3.007 \pm 0.017 \, \MeV$. At much lower space dimensions, the VS-DMRG approach thus yields much lower uncertainties compared to CI ($\Etwop = 3.141 \pm 0.205 \, \MeV$).
For a given size of the many-body space the MPS wavefunction includes correlations much more efficiently compared to CI.

\begin{figure*}[t]
    \centering
    \includegraphics[clip=,width=0.95\textwidth]{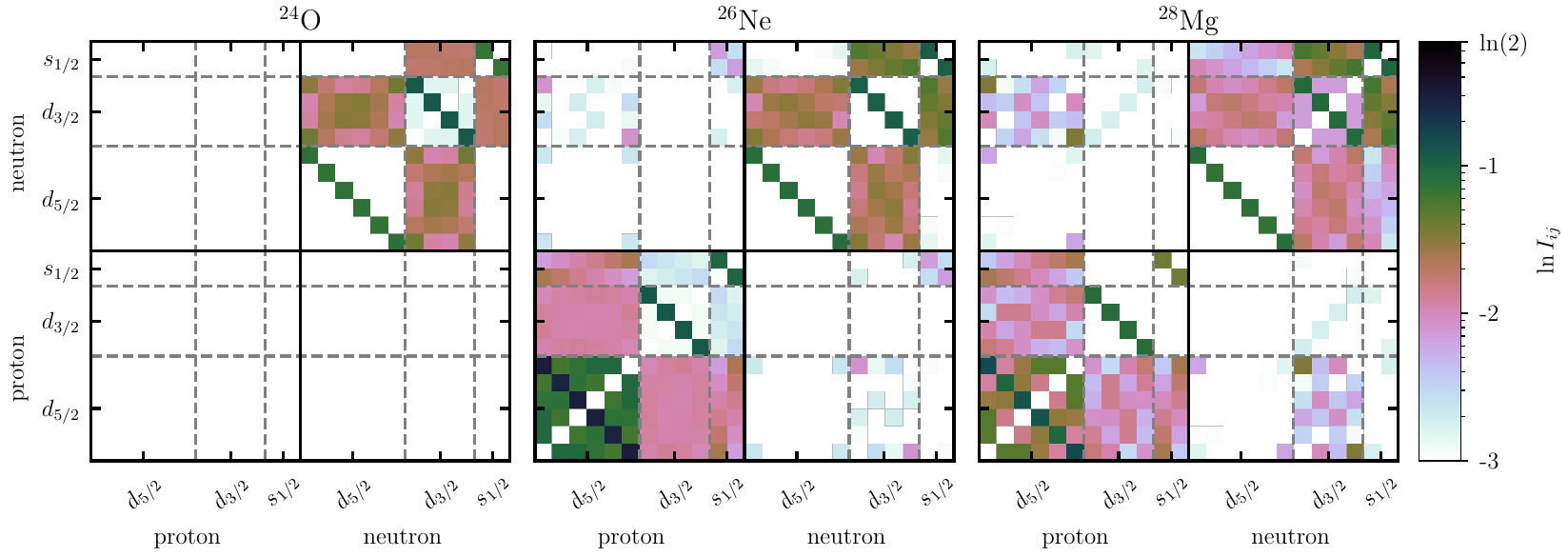}
    \caption{Logarithm of mutual information, $\ln I_{ij}$, for \elem{O}{24}, \elem{Ne}{26}, and \elem{Mg}{28} obtained from VS-DMRG calculations in the $sd$-shell valence space.}
    \label{fig:MIONeMg}
\end{figure*}

Next we study the emergence of shell structure from the perspective of the information entropy from our VS-DMRG calculations.
Figure~\ref{fig:nickelshell} displays neutron, proton and total entropies and $2^+$ excitation energies for \elem{Ni}{70 - 80}.
The total entropy shows a pronounced kink for \elem{Ni}{78} consistent with its doubly magic nature.
The proton contribution to the total entropy is small from \elem{Ni}{70} to \elem{Ni}{78} and then exhibits a strong increase to \elem{Ni}{80}. We attribute this sudden increase of proton correlations to the onset of nuclear deformation effects. This is also consistent with the rapid transition from spherical to deformed ground states beyond \elem{Ni}{78} predicted in Ref.~\cite{Taniuchi2019}.
As expected from the VS-IMSRG results in Ref.~\cite{Taniuchi2019}, the VS-DMRG reproduces nicely the high 2$^+$ excitation energy in \elem{Ni}{78}, with an improved result of $\Etwop = 3.01\, \MeV$ compared to the published VS-IMSRG excitation energy $\Etwop \lesssim 3.34 \, \MeV$. The difference
to the experimental value of $\Etwop = 2.6 \, \MeV$ is therefore significantly decreased for this 1.8/2.0 NN+3N Hamiltonian, and the difference is attributed
to truncated three-body operators in the VS-IMSRG~\cite{Taniuchi2019,Hage16Ni78}.
Finally, we note that the convergence with increasing bond dimension $M$ is significantly slower in \elem{Ni}{78}, which is consistent with the importance of higher $n$-particle-$n$-hole (\ph{n}) correlations in the ground and excited states (see also Fig.~\ref{fig:ni78conv}).

\section{Shell structure in $sd$-shell nuclei}
Following $^{78}$Ni, we explore shell structure in the $sd$ shell based on the total entropies obtained from VS-DMRG calculations using the VS-IMSRG decoupled Hamiltonian from the same 1.8/2.0 NN+3N interactions.
Figure~\ref{fig:ItotONeMg} shows the total neutron and proton entropies for the oxygen isotopes and the $N=16$ isotones \elem{Ne}{26} $(Z=10)$ and \elem{Mg}{28} ($Z=12$). 
Since an $sd$-shell valence space is employed, the proton entropy for the oxygen isotopes is identically zero in all cases.
For the even-mass oxygen isotopes one observes a pronounced kink in the single-orbital entropy at $N=16$, indicating the strong shell closure for \elem{O}{24}.
A complementary analysis of the CI coefficients reveals that the ground state is dominated by the reference state $(\approx 92\%)$ with admixtures from \ph{2}-excitations $(\approx 7\%)$, thus confirming the weakly correlated nature of the many-body state.
A less pronounced kink is observed in \elem{O}{22} where the $d_{5/2}$ shell is closed.
For odd-mass nuclei the entropy is lower compared to their neighbors with an additional neutron due to the presence of an unpaired nucleon.
Note that the entropy of odd-mass nuclei depends on the particular value of the magnetic quantum number $M_J$ in the ground-state multiplet~\cite{Kruppa2022}. Here we consistently show the entropy values for $M_J=1/2$, but differences  for different $M_J$ are small, $\Delta I_\text{tot} \approx 0.1$, and thus do not affect our general conclusions.
Finally, we note that the neutron entropy for \elem{O}{27,28} vanishes due to the single Slater-determinant ground state in the $sd$ shell.

The correlations of \elem{Ne}{26} and \elem{Mg}{28} both reveal an enhancement of the neutron total entropy induced by the presence of valence protons (Fig.~\ref{fig:ItotONeMg}, right panel). 
Both nuclei admit for more collective many-body states with enhanced mixing from \ph{3} excitations (10\%, 17\% in \elem{Ne}{26}, \elem{Mg}{28}, respectively) and \ph{4} excitations (12\%, 15\%).
Deformation effects present in neon and magnesium isotopes cannot be captured within a $sd$-shell valence space but require the inclusion for several major shells~\cite{Tsunoda2017sdpf,Hagen2022PCC}. 
However, this poses challenges in the VS-IMSRG decoupling which is beyond the scope of the present paper and left for future studies~\cite{Miyagi2020}.

A refined understanding of the individual correlation effects is obtained from the mutual information (MI). Figure~\ref{fig:MIONeMg} shows the MI of the $sd$-shell orbitals for \elem{O}{24}, \elem{Ne}{26}, and \elem{Mg}{28}.
In the case of even-mass nuclei with $J^\pi = 0^+$ ground states, the MI for the different $m_j$ orbital substates are degenerate.
The large diagonal entries (black regions) in the proton-proton and neutron-neutron subblock reflect pairing correlations between time-reversed single-particle states~\cite{Legeza2015}.
In \elem{O}{24}, the homogeneous strength in the neutron-neutron blocks $d_{5/2}$-$d_{3/2}$ and $s_{1/2}$-$d_{3/2}$, as well as the uniform MI background in the $d_{3/2}$-$d_{3/2}$ blocks can be understood in terms of nucleon pair fluctuations in generalized seniority-like states~\cite{Kruppa2021}. The proton-proton block of the MI in \elem{Ne}{26} can be similarly understood, and is very similar to the neutron-neutron-block in \elem{O}{18} (not shown).
The emerging structures in the proton-neutron blocks in \elem{Ne}{26} and \elem{Mg}{28} share common features, e.g., the formation of neutron-proton pairs built from $m_j = \pm 5/2$ states.
Moreover, both nuclei admit for enhanced couplings between neutron $d_{3/2}$ and proton $d_{5/2}$ states.
Similar pairing correlations were observed in recent no-core studies of \elem{He}{4,6}~\cite{Robin2021}.

\section{Conclusion and outlook}
In this Letter we performed the first \ai{} DRMG calculations of medium-mass nuclei based on chiral NN+3N interactions.
Combining the DMRG with the VS-IMSRG leads to a powerful hybrid many-body approach, the VS-DMRG, that efficiently accounts for static and dynamic correlation effects. 
The use of an MPS parametrization of the many-body state is computationally superior to conventional CI expansions, and enables convergence in large-scale valence-space applications.
As shown for \elem{Ni}{78} and in the $sd$ shell, the VS-DMRG through its entropy-based entanglement measures also provides new insights to shell structure and correlations in nuclei.
Moreover, the VS-DMRG is ideally suited for exploring systems that are not captured starting from a single-reference state, such as deformed nuclei.
However, this requires the use of multi-shell decoupling in the VS-IMSRG which is still an open area~\cite{Miyagi2020}.
While the present focus was on the calculation of energies, the VS-DMRG framework can be naturally extended to other observables such as radii or electroweak transitions.
For future developments, the use of a symmetry-restricted, i.e., $J$-scheme, formulation of the VS-DMRG (see, e.g., Refs.~\cite{Pittel2006,Thakur2008jdmrg}) will be helpful to cope with the increasing number of orbitals in large-scale applications.
Furthermore, the study of multi-partite entanglement~\cite{Szalay-2017,Brandejs2019} can provide insights to many-body correlations in nuclei.

\section*{Acknowledgements}

We thank J. Hoppe for providing the VS-IMSRG matrix elements for the $sd$-shell calculations, and J. Men\'endez and J. Holt for sharing the VS-IMSRG matrix elements used in Ref.~\cite{Taniuchi2019}.
The work of A.T. and A.S. was supported by the European Research Council (ERC) under the European Union's Horizon 2020 research and innovation programme (Grant Agreement No.~101020842).
\"O.L. and G.Z. have been supported by the Quantum Information National Laboratory of Hungary and by the Hungarian National Research Development and Innovation Office (NKFIH) through Grants Nos.~K120569, K134983, TKP2021-NVA and SNN139581. \"O.L. received further support from the Hans Fischer Senior Fellowship programme funded by the Technical University of Munich -- Institute for Advanced Study and from the Center for Scalable and Predictive methods for Excitation and Correlated phenomena (SPEC), funded as part of the Computational Chemical Sciences Program by the U.S.~Department of Energy (DOE), Office of Science, Office of Basic Energy Sciences, Division of Chemical Sciences, Geosciences, and Biosciences at Pacific Northwest National Laboratory.
C.P.M. was supported by UEFISCDI under project No.~PN-III-P4-ID-PCE-2020-0277, 
and the project for funding the excellence, contract No.~29 PFE/30.12.2021.
M.A.W. was supported by the Janos Bolyai Research Scholarship of the Hungarian Academy of Sciences and by the UNKP-22-4 and UNKP-22-5-BME-330 New National Excellence Program of the Ministry for Culture and Innovation from the source of the National Research, Development and Innovation Fund.
\bibliographystyle{apsrev4-1}
\bibliography{strongint}

\end{document}